\documentclass{PoS}
\usepackage{amsmath,graphicx}

\title{From C to Parton Sea: Bjorken-x Dependence of the PDFs}
\ShortTitle{From C to Parton Sea: Bjorken-x Dependence of the PDFs}

\author{\speaker{Huey-Wen Lin}
\\
Department of Physics and Astronomy, Michigan State University, East Lansing, MI 48824 \\
Department of Computational Mathematics,
  Science and Engineering, Michigan State University, East Lansing, MI 48824
}

\abstract{
Studying the structure of nucleons is not only important to understanding the strong interactions of quarks and gluons, but also to improving the precision of new-physics searches. Since a broad class of experiments, including the LHC and dark-matter detection, require Standard-Model backgrounds with parton distribution functions (PDFs) as inputs for disentangling SM contributions from potential new physics. For a long time, lattice calculations of the PDFs (as well as many hadron structures) has been limited to the first few moments. In this talk, we present a first direct calculation of the Bjorken-x dependence of the PDFs using Large-Momentum Effective Theory (LaMET). An exploratory study of the antiquark/sea flavor asymmetry of these distributions will be discussed. This breakthrough opens an exciting new frontier calculating more complicated quantities, such as gluon structure and transverse-momentum dependence, which will complement existing theoretical programs for the upcoming Electron-Ion Collider (EIC) or Large Hadron-Electron Collider (LHeC).
}

\FullConference{34th annual International Symposium on Lattice Field Theory\\
		24-30 July 2016\\
		University of Southampton, UK}

\begin{document}

\section{Introduction}

\subsection{Parton Distribution Functions}
\begin{figure}
\begin{center}
\includegraphics[width=0.6\textwidth]{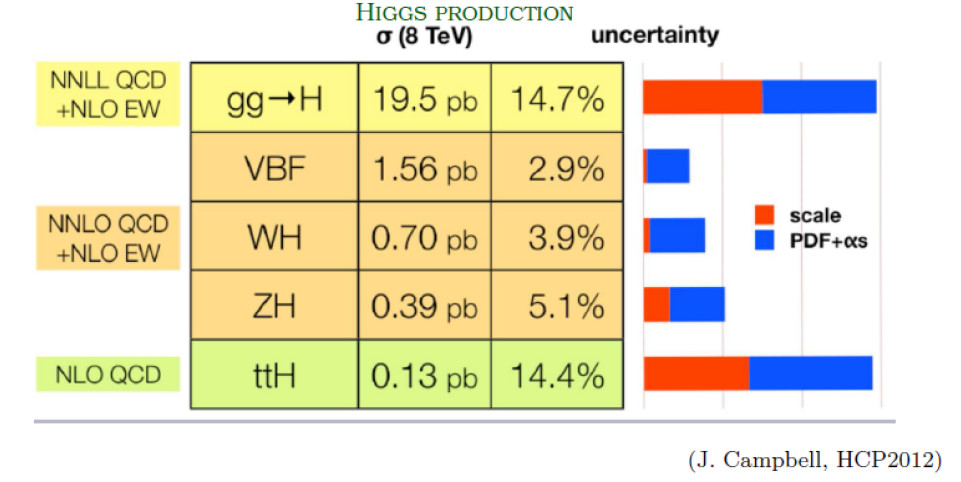}
\caption{Uncertainty breakdown of Higgs production in various channels. The uncertainty due to $\alpha_s$ and PDFs dominates many Higgs-production channels. 
}
\label{fig:HiggsCS}
\end{center}
\end{figure}

Parton distribution functions (PDFs) provide a universal description of hadronic constituents as well as 
critical inputs for the discovery of the Higgs boson found at the Large Hadron Collider (LHC)
through proton-proton collisions~\cite{CMS:2012nga,ATLAS:2012oga}. Despite 
this great discovery, the LHC has many tasks remaining,
and the focus of the future Runs 2--5 will be to search for physics beyond the
Standard Model (BSM). 
In order to probe new physics at the LHC, we need to improve the precision with which we
know the Standard-Model (SM) backgrounds such that one can discriminate between these and
new-physics signatures. Unfortunately, our knowledge of
many  
cross sections remains dominated by PDF uncertainties~\cite{Butterworth:2015oua,Badger:2016bpw,Heinemeyer:2013tqa}; 
for example, Fig.~\ref{fig:HiggsCS} shows a few Higgs-production channels having PDF-dominated uncertainties.  
Thus, improvement of current PDF uncertainties is important to assist new-physics searches during the ongoing 14-TeV collisions at LHC
and for future hadron colliders.

In addition to their applications to the energy frontier, PDFs also reveal
nontrivial structure inside the nucleon, such as the momentum and spin distributions 
of partons. Many ongoing and planned experiments at facilities around 
the world, such as Brookhaven and Jefferson Laboratory in the United States, GSI in Germany, J-PARC in
Japan, or a future Electron-Ion Collider (EIC) or Large Hadron-Electron Collider (LHeC), are set to explore the less-known 
kinematic regions of the nucleon structure and more. 

Multiple global PDFs analyses have integrated all available experimental data from the past half century, some including the LHC's Run-1 data, 
trying to tease out the best possible understanding of the PDFs. A few recent updates are
CT14~\cite{Dulat:2015mca},
CJ15~\cite{Accardi:2016qay}, 
NNPDF2.3~\cite{Ball:2012cx},
HERAPDF1.5~\cite{CooperSarkar:2011aa},
ABM11~\cite{Alekhin:2012ig} and
MSTW2008~\cite{Martin:2009iq}.
Different groups use different experimental data inputs, parametrizations and assumptions. In cases 
where data is abundant, the derived PDFs agree well; however, when the data are limited or carry large uncertainties,
discrepancies appear. Sometimes the PDF uncertainty is estimated through combined analysis of different PDF sets. 
In the case of heavier quarks, such as strange and charm, 
one often needs to use nuclear data, such as neutrino scattering off heavy nuclei, and the current understanding
of nuclear-medium corrections in these cases is limited.
For the case of the strange
distributions, the uncertainty remains large compared to the precision needed for upcoming LHC data. 
In some cases, the assumption $\overline{s}(x)=s(x)$ made in global analyses can agree with the data due to the 
large uncertainty~\cite{Dulat:2015mca,Accardi:2016qay}. 
One can also take advantage of the $W+c$ associated-production channel to extract strangeness,
but their results are rather puzzling. For example, ATLAS gets 
$(s+\overline{s})/(2\overline{d})=0.96^{+0.26}_{-0.30}$ at $Q^2=1.9\mbox{ GeV}^2$ and 
$x=0.23$~\cite{Aad:2014xca}. CMS performs a global analysis with deep-inelastic scattering (DIS) data and the muon-charge asymmetry in 
$W$ production at the LHC to extract the ratios of the total integral of
strange and anti-strange to the sum of the anti-up and -down, at $Q^2=20\mbox{ GeV}^2$
finding it to be $0.52^{+0.18}_{-0.15}$~\cite{Chatrchyan:2013mza}. 
Future high-luminosity studies may help to improve our knowledge of the strangeness. 
In the polarized case, SU(3)-flavor symmetry is often assumed due to lack of precision 
experimental data. We learn from the unpolarized case that this assumption introduces underestimated uncertainty. 
For charm PDFs, there is significant debate over the magnitude of the intrinsic charm contribution~\cite{Brodsky:1980pb,Brodsky:1981se,Brodsky:2015fna,Jimenez-Delgado:2014zga,Brodsky:2015uwa}. This is mainly due to the data being unable to discriminate between various proposed QCD models or to pin down the size of intrinsic charm. Such deficiencies will become increasingly problematic as the LHC data become more precise. 
We should seek to resolve these questions in the upcoming years.  

\subsection{Lattice QCD}
A nonperturbative approach using first principles, such as lattice QCD (LQCD), provides hope to resolve many of the 
outstanding theoretical disagreements and provide information in regions that are unknown or difficult to observe 
in experiments. LQCD is a regularization of continuum QCD using a discretized four-dimensional spacetime; it contains a small number of natural parameters, such as the strong coupling constant and quark masses. Unlike continuum QCD, LQCD works in Euclidean spacetime (rather than Minkowski), 
and the coupling and quark masses can be set differently than those in our universe.  
The theory contains two scales that are absent in continuum QCD, one ultraviolet (the lattice spacing $a$) and one infrared (the spatial extent of the box $L$); this setup keeps the number of degrees of freedom finite so that LQCD can be solved on a computer. 
For observables that have a well-defined operator in the Euclidean path integral for numerical integration, we can find their values in continuum QCD by taking the limits $a \rightarrow 0$, $L \rightarrow \infty$ and $m_q \rightarrow m_q^\text{phys}$. LQCD is a natural tool to study the structure of hadrons, such as PDFs, since quarks and gluons are the fundamental degrees of freedom.

However, probing hadron structure with lattice QCD has been limited to only the first few moments (integrals over the distributions) for decades, due to complications arising from the breaking of rotational symmetry by the discretized Euclidean spacetime. 
In principle, this problem can be avoided by working with moments of parton distributions, which correspond to matrix elements of local operators, provided all the moments can be computed to recover the whole PDF. In
practice, one can only obtain the first few (about 3) moments due to operator mixing with lower-dimension operators with coefficients proportional to inverse powers of the lattice spacing, which divergent in the continuum limit.
Even if one can design more complicated operators to subtract the power divergence arising from the mixing of high-moment operators to get to even higher moments, the renormalization for the higher-moment operators becomes significantly more complicated,
and the correlators suffer from signal-to-noise problems as well. 

In recent years more and more LQCD nucleon matrix elements have been directly 
calculated at the physical quark mass, a big breakthrough compared with a few years ago. 
Still, the calculations were limited to the first couple leading moments. Higher moments, such as
$\langle x^2 \rangle$, have not been updated using dynamical fermions for more than a decade~\cite{Gockeler:2004wp}. 
There are interesting proposals to obtain higher moments by using smeared sources to overcome the power-divergent mixing problem~\cite{Davoudi:2012ya} and 
by using light-quark--to--heavy-quark transition currents to compute current-current correlators in Euclidean space~\cite{Detmold:2005gg}.
There are also ideas about obtaining the structure functions directly from the hadronic tensor current~\cite{Liu:1993cv,Liu:1998um,Liu:1999ak,Liu:2016djw}. However, none of the above 
ideas have been carried out due to their complexity in the lattice numerical calculation. 

\subsection{New Directions}
A recent paper by X.-D.~Ji~\cite{Ji:2013dva} proposed a much more straightforward way of calculating the full Bjorken-$x$ dependence of PDFs, generalized parton distributions (GPDs) and other quantities without dealing with many moments nor requiring enormous computational resources to achieve. In this large-momentum effective theory (LaMET) framework, 
we take an operator containing an integral of gluonic field strength along a line and boost the nucleon momentum toward the speed of light, tilting the spacelike line segment toward the lightcone direction. The time-independent, nonlocal (in space) correlators at finite $P_z$ can be directly evaluated on the lattice. For example, the quark unpolarized distribution can be calculated via
\begin{equation}
q_\text{lat}(x,\mu,P_z)= \int \frac{dz}{4\pi} e^{izk} \times 
  \left\langle \vec{P} \right| \bar{\psi}(z) \gamma_z
      \left( \prod_n U_z(n\hat{z})\right) \psi(0) \left| \vec{P} \right\rangle,
\end{equation}
where $U_z$ is a discrete gauge link in the $z$ direction, 
$x=k/P_z$, $\mu$ is the renormalization scale 
and $\vec{P}=\{0,0,P_z\}$ is the momentum of the hadron, 
taken such that $P_z\rightarrow\infty$. Since no amount of boost will take the nucleon exactly onto the lightcone, there remain corrections power-suppressed by $P_z$ as 
$O\left({M_H^2}/{P_z^2},{\Lambda^2_\text{QCD}}/{P_z^2}\right)$ (where $M_H$ is the hadron mass). 
The same idea can be straightforwardly applied to the helicity $\Delta q(x,\mu)$ and transversity $\delta q(x,\mu)$ distributions
for the direct lattice-QCD calculation of these quantities~\cite{Lin:2014zya,Chen:2016utp}. 

The above distribution is what we call the ``quasi-distribution'' whose shape depends on $P_z$, the gauge-links used, and other parameters.
Therefore, improvements are needed to recover the true lightcone distribution. 
First, after renormalization, one needs to match the boosted momentum $P_z \rightarrow \infty$ in the continuum limit; 
the factor $Z\left(\xi=\frac{x}{y},\frac{\mu}{P_z},\frac{\Lambda}{P_z}\right)$
has been computed up to one loop in Ref.~\cite{Xiong:2013bka} for the non-singlet cases. 
Secondly, the leading mass-correction comes at $(M_N/P_z)^2$ order, which can be significant for the 
nucleon case; see our previous work~\cite{Lin:2014zya,Chen:2016utp} for the explicit formulation for these.
Lastly, there are also improvements needed related to higher-twist operators, 
$(\Lambda_\text{QCD}/P_z)^2$ (for details see Ref.~\cite{Chen:2016utp}); we remove these through $1/P_z^2$ extrapolation. We have implemented each of these improvements in our pioneering
results on the calculation of these distributions on the lattice, shown in Sec.~\ref{sec:preWorks}. 

\section{Previous Exploratory Lattice Study}\label{sec:preWorks}

This section reports the results of the first round of exploratory studies using the LaMET method, which have demonstrated
the success of the approach. 
The LaMET approach for direct calculation of the $x$-dependence of parton distributions was first implemented on the lattice by the PI of this proposal, and was first reported~\cite{Lin:2014gaa,Lin:2014yra} at various conferences in the summer of 2013.
This was done using $N_f=2+1+1$ (up/down, strange and charm loops in the QCD vacuum) highly improved staggered quarks (HISQ) lattice gauge ensembles (generated by the MILC Collaboration) and
clover valence fermions with pion mass 310~MeV at coarse lattice spacing 0.12~fm.
Preliminary studies of the Bjorken-$x$ dependence of the quark, helicity and transversity distributions show reasonable signals for the quasi-distributions. 
In 2014, we reported the first attempt to make a lattice calculation of the unpolarized and longitudinally polarized isovector quark distributions using the LaMET formalism~\cite{Lin:2014zya} including the one-loop and mass corrections.  
Earlier this year, we updated the mass-correction formulation, and refined the analysis of the polarized distributions~\cite{Chen:2016utp}.  
Below we give a summary of the steps needed for the calculations, and discuss highlights of the results of the exploratory study. 


On the lattice, we first calculate the time-independent, nonlocal (in space, chosen to be the $z$ direction) correlators of a nucleon with finite-$P_z$ boost
\begin{align}
\label{eq:hlat}
h_\text{lat}^\Gamma(z,\mu,P_z) =  
  \left\langle \vec{P} \right|
    \bar{\psi}(z) \Gamma \left( \prod_n U_z(n\hat{z})\right) \psi(0)
  \left| \vec{P} \right\rangle,
\end{align}
where 
$\Gamma=\gamma_z$, $\gamma_z\gamma_5$ or $\gamma_z\gamma_\perp \gamma_5$ for the quark density (unpolarized), helicity (longitudinally polarized) or transversity (transversely polarized) distributions, respectively. 
Figure~\ref{fig:HiggsCS}
illustrates the diagrams involved for the three-point correlator with examples of ``connected'' and ``disconnected'' contractions.
In the exploratory study, we have focused on isovector quantities (mainly the up/down flavor asymmetry), where the expensive ``disconnected'' diagrams are canceled. 
To account for excited-state contamination, the calculations are done using two source-sink nucleon separations, 0.96 and 1.2~fm, and the ground-state signal is extracted using a simultaneous two-state fit of the nucleon matrix-element correlators; 
the detailed procedure is described in Ref.~\cite{Bhattacharya:2013ehc} for the nucleon charges.
Examining the individual fits to each source-sink nucleon separation, we do not see noticeable excited-state 
contamination for either separation within statistical errors. 
The bare lattice nucleon matrix elements are calculated with three boost momenta: $P_z=\{1,2,3\} 2\pi/L$, which correspond to nucleon momenta of 0.43, 0.86 and 1.29~GeV, respectively. 
In all three cases (unpolarized, helicity and transversity) studied in Refs.~\cite{Lin:2014zya,Chen:2016utp}, the matrix elements vanish when the link length reaches 10--12. 
The signal-to-noise ratios worsen as the nucleon is increasingly boosted, so to push this method forward, 
future studies would investigate methods for improving nucleon momentum sources.

\begin{figure}
\includegraphics[width=0.32\textwidth]{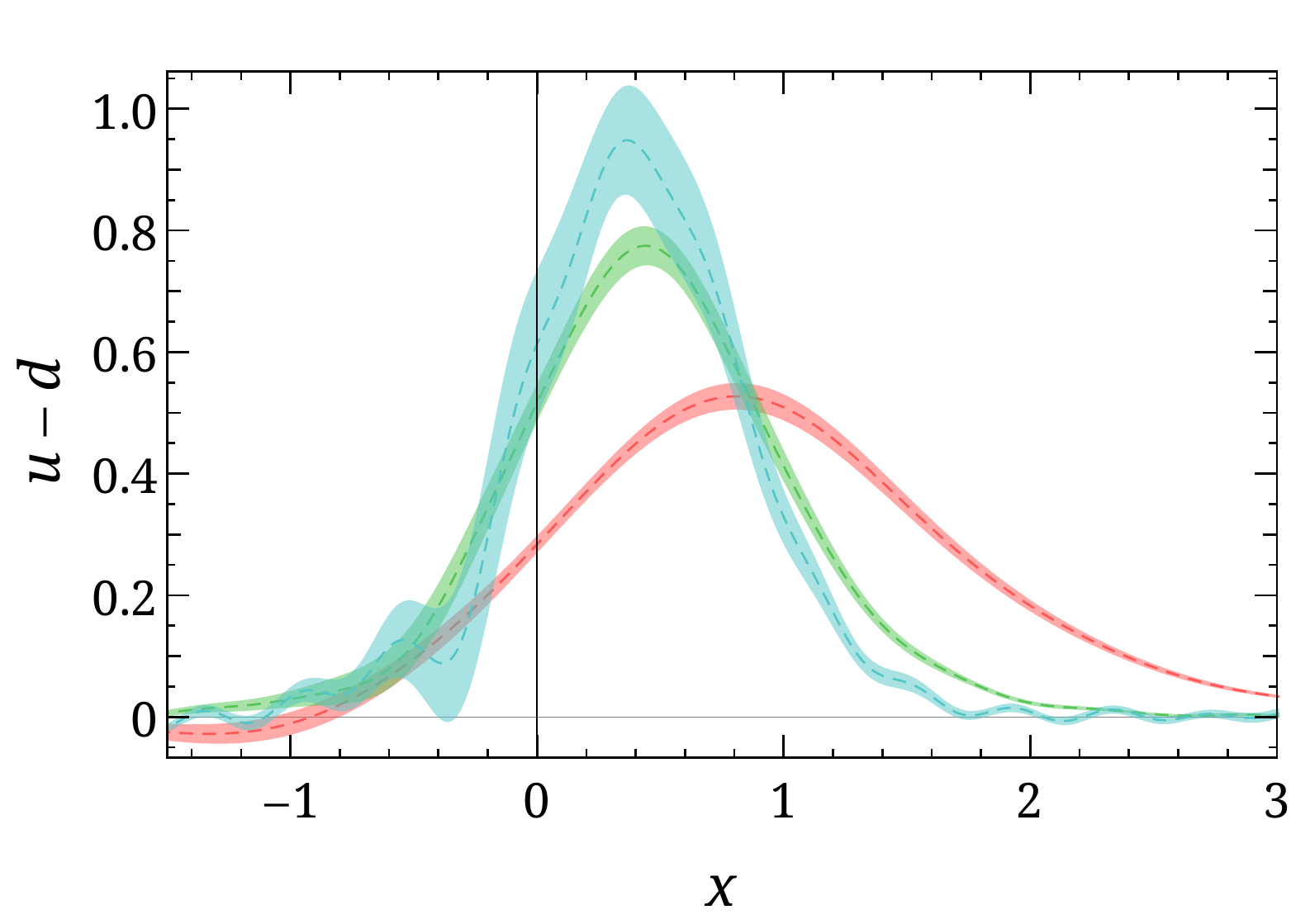}
\includegraphics[width=0.32\textwidth]{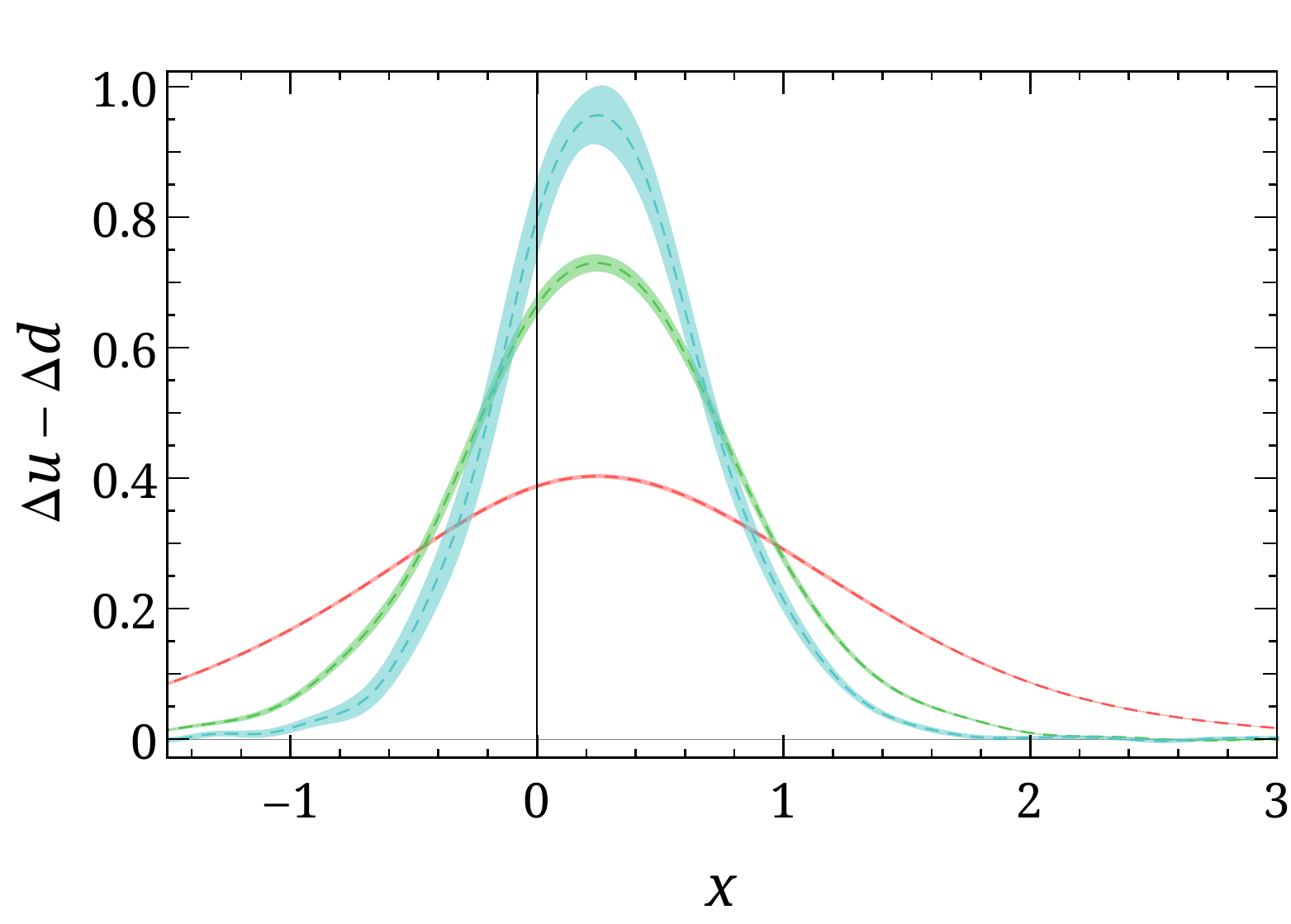}
\includegraphics[width=0.32\textwidth]{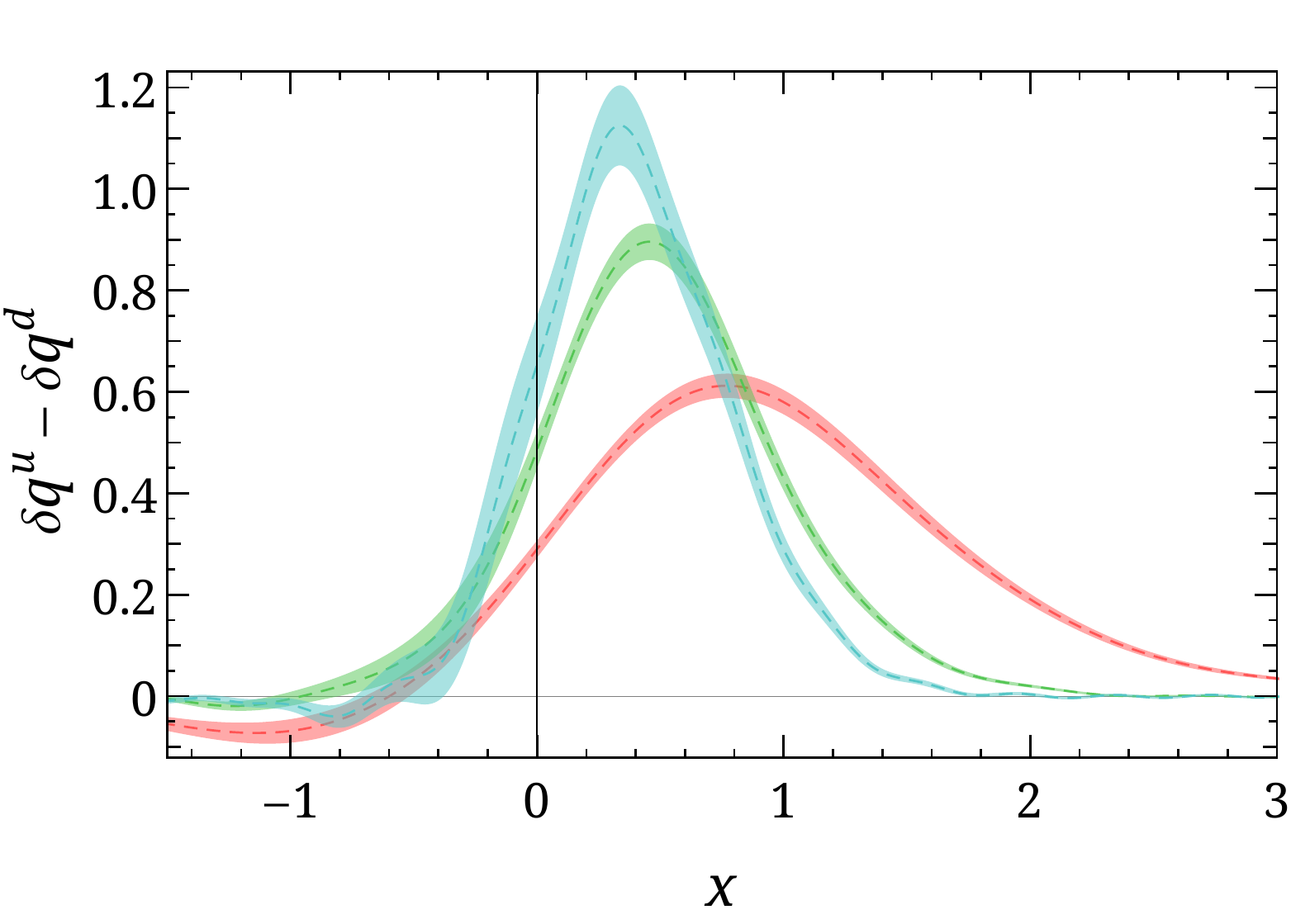}
\caption{
The nucleon isovector quasi-distributions for the parton density (left), helicity (middle) and transversity (right) as functions of $x$. The different colors indicate $P_z$ (in units of $2\pi/L$) 1 (red), 2 (green), 3 (cyan). 
}\label{fig:distribution}
\end{figure}

We then take the integrals to Fourier transform the lattice matrix elements as functions of spatial link length $z$ into the quasi-distributions as functions of parton momentum fraction $x=k/P_z$:
\begin{align}
\label{eq:qlat}
q_\text{lat}^\Gamma(x,\mu,P_z) = \int \frac{dz}{4\pi} e^{izk} \times h_\text{lat}^\Gamma(z,\mu,P_z).
\end{align}
Since the matrix elements go to zero beyond about 12, the integral does not depend sensitively on the choice of maximum $z$ in the range from 10 to 15.   
The normalization of the long-link operators is currently 
estimated through zeroth moment of the quark distribution (assuming the lattice renormalization for $q(x)$ is multiplicative~\cite{Ji:2015jwa}),
\begin{equation}
q_\text{lat}^\Gamma(x,\mu,P_z) = \frac{q_\text{lat}^\Gamma(x,\mu,P_z)}{\int\!dx\,q_\text{lat}^\Gamma(x,\mu,P_z)} \times
  g_\Gamma^\text{local}(\mu=2\mbox{ GeV})_{\overline{\text{MS}}}.
\end{equation}
This choice reduces the systematic uncertainty arising from the matching
and other systematics such as finite-volume effects and lattice discretization. 
Given that the lattice renormalization constants for most observables are close to 1 on this ensemble,
we will get reasonable cancellation of the remaining factors. 
In all three quasi-distributions, shown in Fig.~\ref{fig:distribution}, the smallest momentum has the widest distribution, spreading out to 
large positive and negative $x$, beyond $|x|=1$. But as the boosted momentum increases, the distribution sharpens
and narrows, decreasing the contribution coming from the $|x|>1$ regions, just what we would expect 
in the lightcone distribution. 
This is not hard to understand (as discussed in Ref.~\cite{Lin:2014zya}): in the infinite-momentum frame, no constituents of the nucleon can carry more momentum than the nucleon as a whole. However, since the momentum in our calculation is finite, the PDF does not have to vanish at $x=1$. 

\begin{figure*}
\centering
 \includegraphics[width=0.54\textwidth]{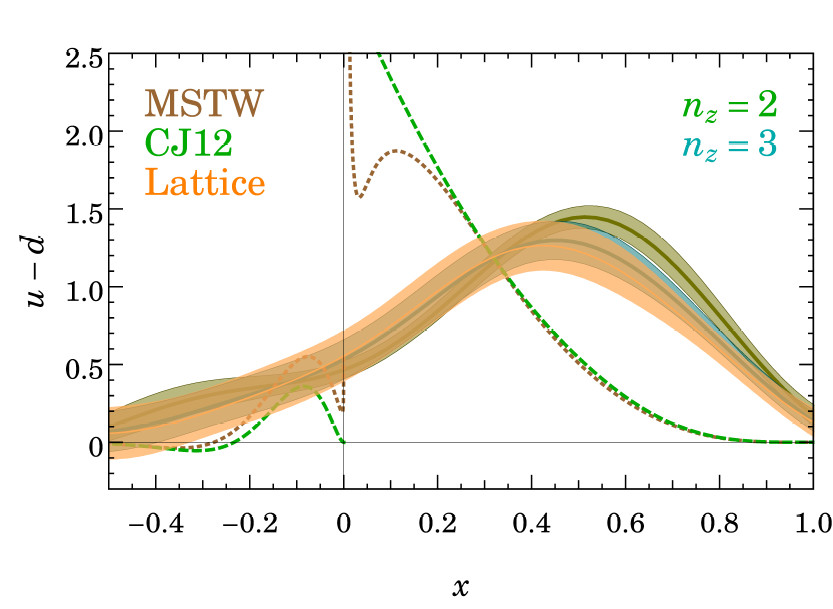}
\caption{
The unpolarized isovector quark distribution $u(x)-d(x)$ computed on
the lattice (orange band: final extrapolation, gold band: $n_z=2$, cyan band: $n_z=3$ with $\vec{P}=\{0,0,n_z\}\frac{2\pi}{L}$), compared with the global analyses by 
MSTW~\cite{Martin:2009iq} (brown dotted line),
and CTEQ-JLab (CJ12, green dashed line)~\cite{Owens:2012bv}.
}
\label{fig:quark-dist1}
\end{figure*}

The resulting quark distribution after one-loop and mass corrections is
shown in Fig.~\ref{fig:quark-dist1}. 
Although the uncorrected distributions have a strong dependence on $P_z$,
the corrected distributions have much reduced $P_z$ dependence. 
We further reduce the remaining leading
$\mathcal{O}(\Lambda_\text{QCD}^2/P_z^2)$ correction by extrapolating to infinite momentum using the
form $a + b /P_z^2$. 
The resulting distribution for the $|x| > 1$ region is within 2 sigma of zero;
thus, we recover the correct support for the physical distribution within error.
Note that the smallest reliable region of $x$ is related to the largest
momentum on available on the lattice $\mathcal{O}(1/a)$, which is roughly the
inverse of length of the lattice volume in the link direction; therefore, 
we expect large uncertainty for $x \in [-0.05, 0.05]$.

The isovector unpolarized 
distribution $u(x)-d(x)$ is shown in Fig.~\ref{fig:quark-dist1} as the orange
band, along with the distributions from the largest two momenta $P_z$ to show convergence.  
Compared with global analyses by CTEQ-JLab (CJ12)~\cite{Owens:2012bv} and NLO
MSTW08~\cite{Martin:2009iq} at $\mu \approx 1.3$~GeV, the
lattice distribution weighs more at larger $\left\vert x\right\vert$. This is
qualitatively consistent with artifacts due to using heavier-than-physical
light-quark masses. The heavier quarks naturally reduce the long-range
contributions to the correlator, 
which after Fourier transformation
reduce the small-$\left\vert x\right\vert$ contribution to the distribution.
Since the total integrated $u-d$ is conserved, a reduction in
small-$\left\vert x\right\vert $ means an increase in
larger-$\left\vert x\right\vert$.
This is also consistent with the fact that the lattice first-moment of the momentum
fraction ($\langle x\rangle_{u-d}$) and helicity ($\langle x\rangle_{\Delta u
-\Delta d}$) above pion mass 250~MeV is roughly double the integrated values
derived from global analyses~\cite{Lin:2009qs,Alexandrou:2013cda}. It would be
very interesting to observe how the distribution changes when approaching the
physical pion mass.

\begin{figure*}
\centering
 \includegraphics[width=0.48\textwidth]{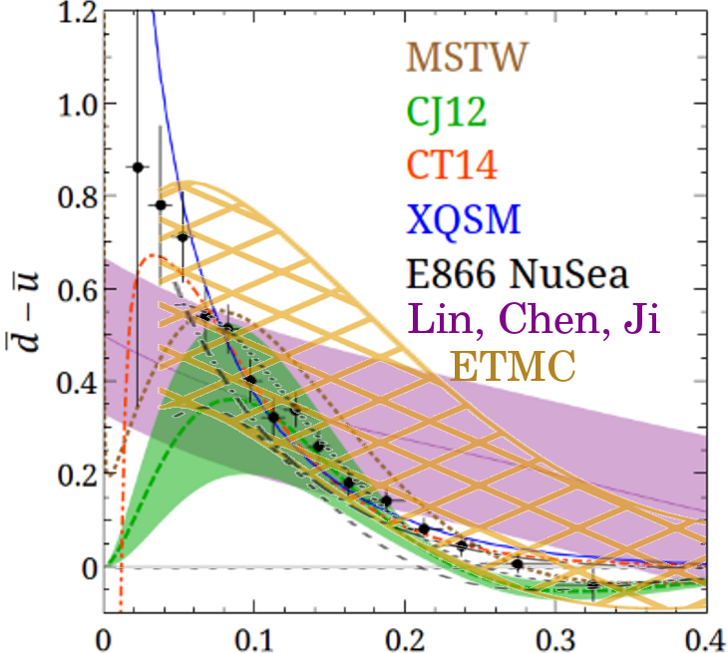}
\caption{
The sea flavor asymmetry measured in the Drell-Yan experiment E866 at FNAL (black circles), along with our 2014 exploratory study at 310-MeV pion mass~\cite{Lin:2014zya} (purple band), 2015 ETMC~\cite{Alexandrou:2015rja} (yellow) and the global analyses by
MSTW~\cite{Martin:2009iq} (brown dotted line),
and CTEQ-JLab (CJ12, green dashed line)~\cite{Owens:2012bv}.
}
\label{fig:quark-dist2}
\end{figure*}

Our pioneering study also allows us access to antiquark structure that could
never have been studied in the traditional lattice approach; see 
Fig.~\ref{fig:quark-dist2} for the sea flavor asymmetry obtained in our
exploratory study~\cite{Lin:2014zya}. 
The sea-quark distribution is obtained from the negative-$x$ contribution:
$\overline{q}(x)=-q(-x)$. Our result favors a large asymmetry in the
distributions of sea up and down antiquarks in the nucleon
with $\int_0^\infty dx\,(\overline{u}(x)-
\overline{d}(x))= 0.13(7)$, which was first observed by the New Muon
Collaboration (NMC) through the cross-section ratio for deep inelastic
scattering of muons from hydrogen and deuterium~\cite{Arneodo:1994sh}, and
later confirmed by other experiments using different processes, such as
Drell-Yan at E665~\cite{Adams:1995sh} and E866/NuSea~\cite {Towell:2001nh}. 
For the first time in LQCD history, we can directly calculate the antiquark asymmetry;
Our result is close
to the experimental one obtained by NMC in their DIS measurement, $0.147(39)$ at
$Q^2=4\mbox{ GeV}^2$ and by HERMES in their semi-inclusive DIS (SIDIS) result,
$0.16(3)$ at $Q^2 = 2.3\mbox{ GeV}^2$~\cite{Ackerstaff:1998sr}. 
In independent follow-up lattice work one year later, our result was confirmed by ETMC Collaboration~\cite{Alexandrou:2015rja} using twisted-mass fermion action. 
The traditional lattice approach using moments would require
knowledge of all moments to isolate the antiquark distribution.
Thus, our result on the antiquark distribution is a clear demonstration that 
 our method reaches beyond previous moment calculations in lattice QCD.
With today's computational resources, such calculations could soon be greatly improved by performing them at the physical pion mass with better systematics control.

\begin{figure*}
\includegraphics[width=0.45\textwidth]{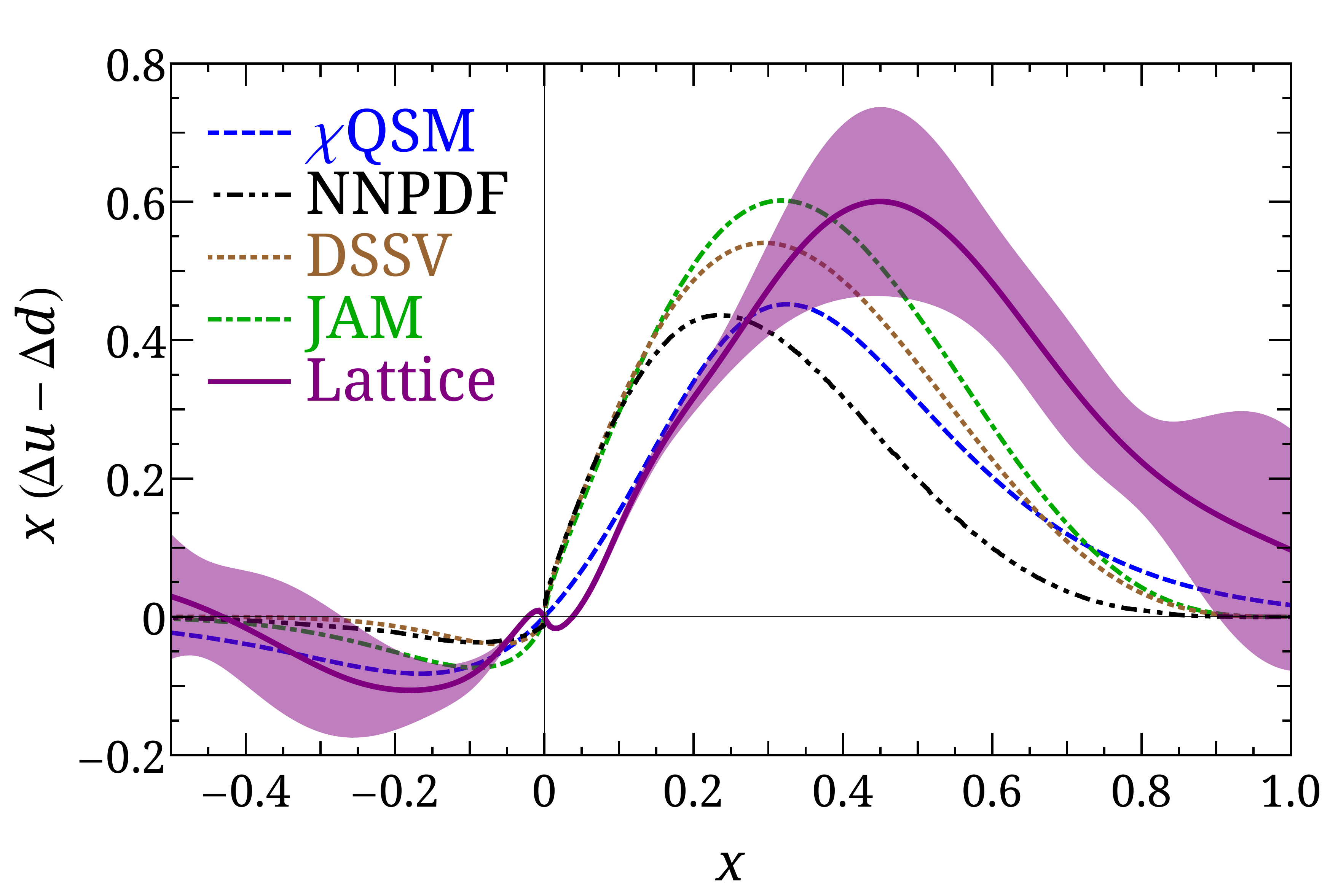}
\includegraphics[width=0.45\textwidth]{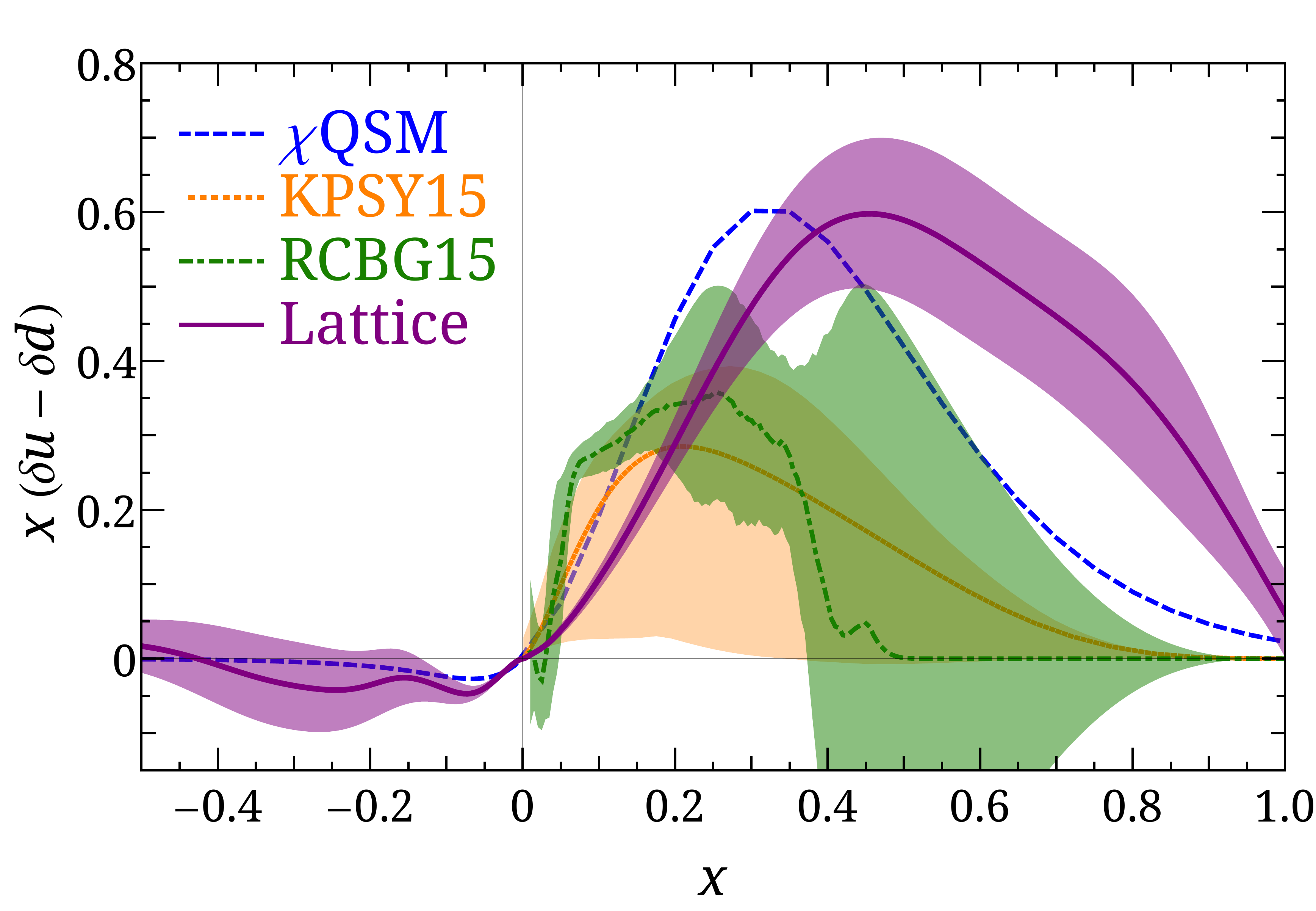}
\caption{
The isovector helicity distribution $x (\Delta u(x)-\Delta d(x))$ (left) and isovector transversity distribution $x(\delta u(x)-\delta d(x))$ (right)
(purple band) computed on the lattice, shown in purple bands. 
The uncertainties in the global analyses are omitted here for visibility reasons, 
and they are NNPDFpol1.1~\cite{Nocera:2014gqa},
JAM~\cite{Jimenez-Delgado:2013boa} (green dot-dashed) and
DSSV09~\cite{deFlorian:2009vb} (brown dotted line),
and a model calculation $\chi$QSM~\cite{Schweitzer:2001sr} (blue dashed line).
For the transversity, we plotted the latest phenomenological analyses from Refs.~\cite{Kang:2015msa} and 
\cite{Radici:2015mwa} (labeled as KPSY15 and RCBG15, orange and green band, respectively).
The corresponding anti-quark distributions are 
$\Delta \overline{q}(x) = \Delta q(-x)$ 
and
$\delta \overline{q}(x) = -\delta q(-x)$.}
\label{fig:helicity-transversity}
\end{figure*}

We show the helicity distribution result from this ensemble in the left panel of Fig.~\ref{fig:helicity-transversity} 
$x (\Delta u(x)-\Delta d(x))$, along with selected recent global analyses
by JAM~\cite{Jimenez-Delgado:2013boa}, DSSV~\cite{deFlorian:2009vb}, and
NNPDFpol1.1~\cite{Nocera:2014gqa}, whose nucleon isovector distribution uncertainties have been ignored here. 
We see more weight distributed in the large-$x$ region, which could shift toward smaller $x$ as we lower 
the quark masses. This is because lower quark mass increases the long-range correlations in ${\Delta h}_\text{lat}(z)$,
which in turn increases the small-$x$ contribution in the Fourier transformation. 
There are noticeable differences between the extracted polarized PDFs
depending on the experimental cuts, theory inputs, parametrization, and so on.
For example, JAM excludes SIDIS data, leaving the sign of the light antiquark
determined by the valence and the magnitude determined from sum rules. 
DSSV also relies on assumptions such as SU(3) symmetry to constrain the 
analysis and adds a very small symmetry-breaking term. A direct lattice study
of hyperon axial couplings~\cite{Lin:2007ap} suggested that SU(3) breaking is
roughly 20\% at the physical point, bigger than these assumptions. 
Similar assumptions are also made by NNPDFpol1.1~\cite{Nocera:2014gqa}.
These assumptions are unavoidable due to the difficulties of getting constraint data from polarized experiments.
Future experiments with neutral- and
charged-current deep inelastic scattering  (such as at EIC) will provide useful measurements
to constrain our understanding of the antiquark helicity distribution.

Our result for antiquark helicity favors more polarized up quark than down
flavor, with a moderate polarized total sea asymmetry,
$\int_{0.08}^1 \Delta \overline{u}(x)-\Delta \overline{d}(x)=0.14(9)$. 
This was first pointed out in our 2014 paper~\cite{Lin:2014zya} 
which concentrated on the sea flavor asymmetry in the unpolarized distribution. 
The sea flavor asymmetry was confirmed  in the full analysis of the RHIC Run-9 data by both
STAR~\cite{Adamczyk:2014xyw} and PHENIX~\cite{Adare:2015gsd} collaborations in the middle-$x$ range, 
but their results do not clarify what the total asymmetry would be. 
$\chi$QSM, a large-$N_c$ model, gives rather different results by predicting a large polarized sea asymmetry: 0.31.
Unfortunately, our current statistical error does not   
help rule out many models yet based on the total sea asymmetry.
On the experimental global analysis side, the total polarized sea asymmetry estimated by DSSV09 is consistent with zero within 2 sigma, and the central value (about 0.07) is also smaller than the unpolarized case. 
The upcoming RHIC data from Run-13 with significantly improved statistics may shed some light on this matter.
The upcoming Fermilab Drell-Yan experiments (E1027/E1039) may also provide precise experimental input on the polarized sea asymmetry magnitude.

The transversity distribution is the least known PDF among the three
studied, because the direct measurements are so challenging to make.
The few attempts to extract the transversity distribution suffer from fundamental defects.
Ref.~\cite{Anselmino:2008jk} makes various assumptions such as the evolution form and that there is no antiquark contribution. 
Radici et~al.~\cite{Radici:2015mwa,Radici:2013vra} use the Soffer inequality and dihadron fragmentation functions with data from HERMES and COMPASS analysis of pion-pair production in DIS off a transversely polarized target for two combinations of ``valence'' ($q+\bar{q}$) helicity distribution and proper $Q^2$ evolution. 
Kang et~al.~\cite{Kang:2015msa} has improved evolutions implemented in their analysis, but they also make the assumption 
that the sea asymmetry is zero. The distribution for positive $x$ goes quickly to zero,
likely due to lack of data.

Our transversity result is shown in the
the right panel of Fig.~\ref{fig:helicity-transversity}, along with an estimate from 
$\chi$QSM~\cite{Schweitzer:2001sr} and the latest transversity fits from Refs.~\cite{Kang:2015msa,Radici:2015mwa}. 
Surprisingly, our result is rather similar to $\chi$QSM within 90\% confidence, but with slower descent to zero in 
the $x\approx 1$ region, similar to the quark distribution. This can be, again, due to the heavier pion mass used in the 
calculation, as well as the need to push for even larger momenta.
In contrast, the phenomenological results from Ref.~\cite{Kang:2015msa} fall faster as $x$ approaches 1, possibly due to the lack of large-$x$ data for constraining the fit. 

Our result favors $\delta \overline{d}(x) > \delta \overline{u}(x)$ with total sea asymmetry
0.10(8), whose central value is still larger than most model predictions (for example, $\chi$QSM estimates $0.082$)
and in contradiction to the assumption that the antiquark is consistent with zero in some
transversity extractions using experimental data~\cite{Anselmino:2008jk,Radici:2013vra,Kang:2014zza,Kang:2015msa}. 
One interesting thing to note is that the central values of the lattice determination of the tensor charge $g_T$ (that is, $\int_{-1}^{+1} dx\, \delta u(x) -\delta d(x)$) extrapolated to the continuum limit from various groups are consistently higher than the phenomenological ones which assume zero total sea asymmetry in transversity; see the summary plot Fig.~10 in Ref.~\cite{Bhattacharya:2015wna}. This may indicate nonzero sea contribution with the same sign as our prediction here, or missing larger-$x$ data in constraining their fit. 
It would be interesting to see whether such a nonzero sea asymmetry remains in the future high-statistics physical quark mass ensemble; it is certainly contrary to traditional expectation. Improved phenomenological analysis with new experimental data would also help to narrow the phenomenological uncertainties and explore the discrepancy.

The cleanest measurement of the transversity would have both a polarized beam and polarized target, 
but such facility does not yet exist; once again, more data are needed.
PHENIX and STAR will be able to help give more insight into this quantity.
Planned experiments, such as SoLID at Jefferson Lab, can provide good transversity measurements for a wide range of positive $x$.
The future EIC would be able to fill in missing data regions.
The Drell-Yan experiment at FNAL (E1027+E1039) can in principle extract
sea-asymmetry information in the near future to settle the size of the total transversely polarized sea.

Our exploratory study has demonstrated that the Bjorken-$x$ dependence of hadron structure can be studied using LQCD through the LaMET approach. We are able to predict the less-known antiquark up/down asymmetry that many ongoing and planned experiments will explore. 
In addition to the application to PDFs,
the LaMET approach can be extended to many other hadron structures such as GPDs, transverse momentum-dependent distributions (TMDs), meson decay amplitudes and more. The lessons we learn studying the PDFs will have big impacts across many hadronic structures for years to come. 
We will advance this innovative new method by investigating its systematics, including
nonperturbative renormalization, lattice discretization, and finite-volume effects. Only when these systematics are under control can LQCD inputs be used in global PDF analyses with the required precision.

There are two key issues that need to be addressed in future improved lattice calculations. 
First, we should improve how the signal and systematics associated with highly boosted hadrons on the lattice.  
Previous works have proposed momentum sources for form factors with large transfer momentum. 
Recent work by RQCD has implemented momentum-smeared sources and demonstrated improvement of signal-to-noise issues in two-point hadrons with high momenta in lattice QCD~\cite{Bali:2016lva}.
See also the poster contribution by B.~Lang in this conference.  
ETMC has also implemented a momentum source and applied it to their calculation, finding significant improvements to their 
results for PDFs~\cite{Alexandrou:2016jqi}.
      
However, when working with such large momenta, we should keep aware of systematics associated with discretization in $(Pa)^n$. In addition, excited-state contamination gets worse at larger momentum. Therefore, better methods for treating excited-state contamination (using multiple source-sink separations or multiple operators that overlapping different states), as well as finer lattice spacing $a$, would be required to get this systematic under control.

Secondly, matching and renormalization issues should be addressed. 
In the current work, we assume the renormalization is multiplicative, by dividing out the integral of the zeroth moment the distribution; this is probably equivalent to a tree-level renormalization procedure.  However, the matching can be $x$-dependent, and we need to address this systematic before we can integrate the lattice PDFs into the global fit. 
Fortunately, there is progress being made. T.~Ishikawa et~al.~\cite{Ishikawa:2016znu} and Chen et~al.~\cite{Chen:2016fxx} have started to address these matching issue in the quasi-distribution approach. Implementing these ideas in the current lattice PDFs calculations will significantly improve the systematics, especially in the small-$x$ region.

In addition to the application to PDF,  LaMET approach can be extended to many other kinds of hadron structure, such as generalized parton distributions (GPDs), transverse momentum-dependent distributions (TMDs), meson decay amplitudes and others. The lessons we learned in studying the PDFs will have a big impact in the wider study of hadronic structures for many years to come.

\section*{Acknowledgments}
The LQCD calculations were performed using the Chroma software
suite~\cite{Edwards:2004sx}.
Computations for this work were carried out in part on facilities of the
USQCD Collaboration, which are funded by the Office of Science of the
U.S. Department of Energy, and on Hyak clusters at the University of Washington
managed by UW Information Technology, using hardware awarded by NSF grant
PHY-09227700. 
This research used resources of the National Energy Research Scientific Computing Center, a DOE Office of Science User Facility supported by the Office of Science of the U.S. Department of Energy under Contract No. DE-AC02-05CH11231.
We thank MILC Collaboration for sharing the lattices used to perform this study.  
HWL thanks the Institute for Nuclear Theory at the University of Washington for its hospitality during the completion of this work. 
The work of HWL was supported in part by the M.~Hildred Blewett Fellowship of the American Physical Society, www.aps.org.

\bibliographystyle{JHEP}

\begin{thebibliography}{10}

\bibitem{CMS:2012nga}
{\bf CMS} Collaboration, S.~Chatrchyan et~al., {\it {A New Boson with a Mass of
  125 GeV Observed with the CMS Experiment at the Large Hadron Collider}},
  {\em Science} {\bf 338} (2012) 1569--1575.

\bibitem{ATLAS:2012oga}
{\bf ATLAS} Collaboration, G.~Aad et~al., {\it {A particle consistent with the
  Higgs Boson observed with the ATLAS Detector at the Large Hadron Collider}},
  {\em Science} {\bf 338} (2012) 1576--1582.

\bibitem{Butterworth:2015oua}
J.~Butterworth et~al., {\it {PDF4LHC recommendations for LHC Run II}},  {\em J.
  Phys.} {\bf G43} (2016) 023001,
  [\href{http://xxx.lanl.gov/abs/1510.0386}{{\tt arXiv:1510.0386}}].

\bibitem{Badger:2016bpw}
J.~R. Andersen et~al., {\it {Les Houches 2015: Physics at TeV Colliders
  Standard Model Working Group Report}},  in {\em {9th Les Houches Workshop on
  Physics at TeV Colliders (PhysTeV 2015) Les Houches, France, June 1-19,
  2015}}, 2016.
\newblock \href{http://xxx.lanl.gov/abs/1605.0469}{{\tt arXiv:1605.0469}}.

\bibitem{Heinemeyer:2013tqa}
{\bf LHC Higgs Cross Section Working Group} Collaboration, J.~R. Andersen
  et~al., {\it {Handbook of LHC Higgs Cross Sections: 3. Higgs Properties}},
  \href{http://xxx.lanl.gov/abs/1307.1347}{{\tt arXiv:1307.1347}}.

\bibitem{Dulat:2015mca}
S.~Dulat, T.-J. Hou, J.~Gao, M.~Guzzi, J.~Huston, P.~Nadolsky, J.~Pumplin,
  C.~Schmidt, D.~Stump, and C.~P. Yuan, {\it {New parton distribution functions
  from a global analysis of quantum chromodynamics}},  {\em Phys. Rev.} {\bf
  D93} (2016), no.~3 033006, [\href{http://xxx.lanl.gov/abs/1506.0744}{{\tt
  arXiv:1506.0744}}].

\bibitem{Accardi:2016qay}
A.~Accardi, L.~T. Brady, W.~Melnitchouk, J.~F. Owens, and N.~Sato, {\it
  {Constraints on large-$x$ parton distributions from new weak boson production
  and deep-inelastic scattering data}},  {\em Phys. Rev.} {\bf D93} (2016),
  no.~11 114017, [\href{http://xxx.lanl.gov/abs/1602.0315}{{\tt
  arXiv:1602.0315}}].

\bibitem{Ball:2012cx}
R.~D. Ball et~al., {\it {Parton distributions with LHC data}},  {\em Nucl.
  Phys.} {\bf B867} (2013) 244--289,
  [\href{http://xxx.lanl.gov/abs/1207.1303}{{\tt arXiv:1207.1303}}].

\bibitem{CooperSarkar:2011aa}
{\bf ZEUS, H1} Collaboration, A.~M. Cooper-Sarkar, {\it {PDF Fits at HERA}},
  {\em PoS} {\bf EPS-HEP2011} (2011) 320,
  [\href{http://xxx.lanl.gov/abs/1112.2107}{{\tt arXiv:1112.2107}}].

\bibitem{Alekhin:2012ig}
S.~Alekhin, J.~Blumlein, and S.~Moch, {\it {Parton Distribution Functions and
  Benchmark Cross Sections at NNLO}},  {\em Phys. Rev.} {\bf D86} (2012)
  054009, [\href{http://xxx.lanl.gov/abs/1202.2281}{{\tt arXiv:1202.2281}}].

\bibitem{Martin:2009iq}
A.~D. Martin, W.~J. Stirling, R.~S. Thorne, and G.~Watt, {\it {Parton
  distributions for the LHC}},  {\em Eur. Phys. J.} {\bf C63} (2009) 189--285,
  [\href{http://xxx.lanl.gov/abs/0901.0002}{{\tt arXiv:0901.0002}}].

\bibitem{Aad:2014xca}
{\bf ATLAS} Collaboration, G.~Aad et~al., {\it {Measurement of the production
  of a $W$ boson in association with a charm quark in $pp$ collisions at
  $\sqrt{s} =$ 7 TeV with the ATLAS detector}},  {\em JHEP} {\bf 05} (2014)
  068, [\href{http://xxx.lanl.gov/abs/1402.6263}{{\tt arXiv:1402.6263}}].

\bibitem{Chatrchyan:2013mza}
{\bf CMS} Collaboration, S.~Chatrchyan et~al., {\it {Measurement of the muon
  charge asymmetry in inclusive $pp \to W+X$ production at $\sqrt s =$ 7 TeV
  and an improved determination of light parton distribution functions}},  {\em
  Phys. Rev.} {\bf D90} (2014), no.~3 032004,
  [\href{http://xxx.lanl.gov/abs/1312.6283}{{\tt arXiv:1312.6283}}].

\bibitem{Brodsky:1980pb}
S.~J. Brodsky, P.~Hoyer, C.~Peterson, and N.~Sakai, {\it {The Intrinsic Charm
  of the Proton}},  {\em Phys. Lett.} {\bf B93} (1980) 451--455.

\bibitem{Brodsky:1981se}
S.~J. Brodsky, C.~Peterson, and N.~Sakai, {\it {Intrinsic Heavy Quark States}},
   {\em Phys. Rev.} {\bf D23} (1981) 2745.

\bibitem{Brodsky:2015fna}
S.~J. Brodsky, A.~Kusina, F.~Lyonnet, I.~Schienbein, H.~Spiesberger, and
  R.~Vogt, {\it {A review of the intrinsic heavy quark content of the
  nucleon}},  {\em Adv. High Energy Phys.} {\bf 2015} (2015) 231547,
  [\href{http://xxx.lanl.gov/abs/1504.0628}{{\tt arXiv:1504.0628}}].

\bibitem{Jimenez-Delgado:2014zga}
P.~Jimenez-Delgado, T.~J. Hobbs, J.~T. Londergan, and W.~Melnitchouk, {\it {New
  limits on intrinsic charm in the nucleon from global analysis of parton
  distributions}},  {\em Phys. Rev. Lett.} {\bf 114} (2015), no.~8 082002,
  [\href{http://xxx.lanl.gov/abs/1408.1708}{{\tt arXiv:1408.1708}}].

\bibitem{Brodsky:2015uwa}
S.~J. Brodsky and S.~Gardner, {\it {Comment on “New Limits on Intrinsic Charm
  in the Nucleon from Global Analysis of Parton Distributions”}},  {\em Phys.
  Rev. Lett.} {\bf 116} (2016), no.~1 019101,
  [\href{http://xxx.lanl.gov/abs/1504.0096}{{\tt arXiv:1504.0096}}].

\bibitem{Gockeler:2004wp}
{\bf QCDSF} Collaboration, M.~Gockeler, R.~Horsley, D.~Pleiter, P.~E.~L. Rakow,
  and G.~Schierholz, {\it {A Lattice determination of moments of unpolarised
  nucleon structure functions using improved Wilson fermions}},  {\em Phys.
  Rev.} {\bf D71} (2005) 114511,
  [\href{http://xxx.lanl.gov/abs/hep-ph/0410187}{{\tt hep-ph/0410187}}].

\bibitem{Davoudi:2012ya}
Z.~Davoudi and M.~J. Savage, {\it {Restoration of Rotational Symmetry in the
  Continuum Limit of Lattice Field Theories}},  {\em Phys. Rev.} {\bf D86}
  (2012) 054505, [\href{http://xxx.lanl.gov/abs/1204.4146}{{\tt
  arXiv:1204.4146}}].

\bibitem{Detmold:2005gg}
W.~Detmold and C.~J.~D. Lin, {\it {Deep-inelastic scattering and the operator
  product expansion in lattice QCD}},  {\em Phys. Rev.} {\bf D73} (2006)
  014501, [\href{http://xxx.lanl.gov/abs/hep-lat/0507007}{{\tt
  hep-lat/0507007}}].

\bibitem{Liu:1993cv}
K.-F. Liu and S.-J. Dong, {\it {Origin of difference between anti-d and anti-u
  partons in the nucleon}},  {\em Phys. Rev. Lett.} {\bf 72} (1994) 1790--1793,
  [\href{http://xxx.lanl.gov/abs/hep-ph/9306299}{{\tt hep-ph/9306299}}].

\bibitem{Liu:1998um}
K.~F. Liu, S.~J. Dong, T.~Draper, D.~Leinweber, J.~H. Sloan, W.~Wilcox, and
  R.~M. Woloshyn, {\it {Valence QCD: Connecting QCD to the quark model}},  {\em
  Phys. Rev.} {\bf D59} (1999) 112001,
  [\href{http://xxx.lanl.gov/abs/hep-ph/9806491}{{\tt hep-ph/9806491}}].

\bibitem{Liu:1999ak}
K.-F. Liu, {\it {Parton degrees of freedom from the path integral formalism}},
  {\em Phys. Rev.} {\bf D62} (2000) 074501,
  [\href{http://xxx.lanl.gov/abs/hep-ph/9910306}{{\tt hep-ph/9910306}}].

\bibitem{Liu:2016djw}
K.-F. Liu, {\it {Parton Distribution Function from the Hadronic Tensor on the
  Lattice}},  {\em PoS} {\bf LATTICE2015} (2016) 115,
  [\href{http://xxx.lanl.gov/abs/1603.0735}{{\tt arXiv:1603.0735}}].

\bibitem{Ji:2013dva}
X.~Ji, {\it {Parton Physics on a Euclidean Lattice}},  {\em Phys. Rev. Lett.}
  {\bf 110} (2013) 262002, [\href{http://xxx.lanl.gov/abs/1305.1539}{{\tt
  arXiv:1305.1539}}].

\bibitem{Lin:2014zya}
H.-W. Lin, J.-W. Chen, S.~D. Cohen, and X.~Ji, {\it {Flavor Structure of the
  Nucleon Sea from Lattice QCD}},  {\em Phys. Rev.} {\bf D91} (2015) 054510,
  [\href{http://xxx.lanl.gov/abs/1402.1462}{{\tt arXiv:1402.1462}}].

\bibitem{Chen:2016utp}
J.-W. Chen, S.~D. Cohen, X.~Ji, H.-W. Lin, and J.-H. Zhang, {\it {Nucleon
  Helicity and Transversity Parton Distributions from Lattice QCD}},  {\em
  Nucl. Phys.} {\bf B911} (2016) 246--273,
  [\href{http://xxx.lanl.gov/abs/1603.0666}{{\tt arXiv:1603.0666}}].

\bibitem{Xiong:2013bka}
X.~Xiong, X.~Ji, J.-H. Zhang, and Y.~Zhao, {\it {One-loop matching for parton
  distributions: Nonsinglet case}},  {\em Phys. Rev.} {\bf D90} (2014), no.~1
  014051, [\href{http://xxx.lanl.gov/abs/1310.7471}{{\tt arXiv:1310.7471}}].

\bibitem{Lin:2014gaa}
H.-W. Lin, {\it {Recent progress on nucleon structure with lattice QCD}},  {\em
  Int. J. Mod. Phys. Conf. Ser.} {\bf 25} (2014) 1460039.

\bibitem{Lin:2014yra}
H.-W. Lin, {\it {Calculating the $x$ Dependence of Hadron Parton Distribution
  Functions}},  {\em PoS} {\bf LATTICE2013} (2014) 293.

\bibitem{Bhattacharya:2013ehc}
T.~Bhattacharya, S.~D. Cohen, R.~Gupta, A.~Joseph, H.-W. Lin, and B.~Yoon, {\it
  {Nucleon Charges and Electromagnetic Form Factors from 2+1+1-Flavor Lattice
  QCD}},  {\em Phys. Rev.} {\bf D89} (2014), no.~9 094502,
  [\href{http://xxx.lanl.gov/abs/1306.5435}{{\tt arXiv:1306.5435}}].

\bibitem{Ji:2015jwa}
X.~Ji and J.-H. Zhang, {\it {Renormalization of quasiparton distribution}},
  {\em Phys. Rev.} {\bf D92} (2015) 034006,
  [\href{http://xxx.lanl.gov/abs/1505.0769}{{\tt arXiv:1505.0769}}].

\bibitem{Owens:2012bv}
J.~F. Owens, A.~Accardi, and W.~Melnitchouk, {\it {Global parton distributions
  with nuclear and finite-$Q^2$ corrections}},  {\em Phys. Rev.} {\bf D87}
  (2013), no.~9 094012, [\href{http://xxx.lanl.gov/abs/1212.1702}{{\tt
  arXiv:1212.1702}}].

\bibitem{Lin:2009qs}
H.-W. Lin, {\it {A Review of Nucleon Spin Calculations in Lattice QCD}},  {\em
  AIP Conf. Proc.} {\bf 1149} (2009) 552--557,
  [\href{http://xxx.lanl.gov/abs/0903.4080}{{\tt arXiv:0903.4080}}].

\bibitem{Alexandrou:2013cda}
C.~Alexandrou, M.~Constantinou, V.~Drach, K.~Hatziyiannakou, K.~Jansen,
  C.~Kallidonis, G.~Koutsou, T.~Leontiou, and A.~Vaquero, {\it {Nucleon
  Structure using lattice QCD}},  {\em Nuovo Cim.} {\bf C036} (2013), no.~05
  111--120, [\href{http://xxx.lanl.gov/abs/1303.6818}{{\tt arXiv:1303.6818}}].

\bibitem{Arneodo:1994sh}
{\bf New Muon} Collaboration, M.~Arneodo et~al., {\it {A Reevaluation of the
  Gottfried sum}},  {\em Phys. Rev.} {\bf D50} (1994) R1--R3.

\bibitem{Adams:1995sh}
{\bf E665} Collaboration, M.~R. Adams et~al., {\it {Extraction of the ratio
  $F2_n / F2_p$ from muon - deuteron and muon - proton scattering at small $x$
  and Q**2}},  {\em Phys. Rev. Lett.} {\bf 75} (1995) 1466--1470.

\bibitem{Towell:2001nh}
{\bf NuSea} Collaboration, R.~S. Towell et~al., {\it {Improved measurement of
  the anti-d / anti-u asymmetry in the nucleon sea}},  {\em Phys. Rev.} {\bf
  D64} (2001) 052002, [\href{http://xxx.lanl.gov/abs/hep-ex/0103030}{{\tt
  hep-ex/0103030}}].

\bibitem{Ackerstaff:1998sr}
{\bf HERMES} Collaboration, K.~Ackerstaff et~al., {\it {The Flavor asymmetry of
  the light quark sea from semiinclusive deep inelastic scattering}},  {\em
  Phys. Rev. Lett.} {\bf 81} (1998) 5519--5523,
  [\href{http://xxx.lanl.gov/abs/hep-ex/9807013}{{\tt hep-ex/9807013}}].

\bibitem{Alexandrou:2015rja}
C.~Alexandrou, K.~Cichy, V.~Drach, E.~Garcia-Ramos, K.~Hadjiyiannakou,
  K.~Jansen, F.~Steffens, and C.~Wiese, {\it {Lattice calculation of parton
  distributions}},  {\em Phys. Rev.} {\bf D92} (2015) 014502,
  [\href{http://xxx.lanl.gov/abs/1504.0745}{{\tt arXiv:1504.0745}}].

\bibitem{Jimenez-Delgado:2013boa}
P.~Jimenez-Delgado, A.~Accardi, and W.~Melnitchouk, {\it {Impact of hadronic
  and nuclear corrections on global analysis of spin-dependent parton
  distributions}},  {\em Phys. Rev.} {\bf D89} (2014), no.~3 034025,
  [\href{http://xxx.lanl.gov/abs/1310.3734}{{\tt arXiv:1310.3734}}].

\bibitem{deFlorian:2009vb}
D.~de~Florian, R.~Sassot, M.~Stratmann, and W.~Vogelsang, {\it {Extraction of
  Spin-Dependent Parton Densities and Their Uncertainties}},  {\em Phys. Rev.}
  {\bf D80} (2009) 034030, [\href{http://xxx.lanl.gov/abs/0904.3821}{{\tt
  arXiv:0904.3821}}].

\bibitem{Nocera:2014gqa}
{\bf NNPDF} Collaboration, E.~R. Nocera, R.~D. Ball, S.~Forte, G.~Ridolfi, and
  J.~Rojo, {\it {A first unbiased global determination of polarized PDFs and
  their uncertainties}},  {\em Nucl. Phys.} {\bf B887} (2014) 276--308,
  [\href{http://xxx.lanl.gov/abs/1406.5539}{{\tt arXiv:1406.5539}}].

\bibitem{Lin:2007ap}
H.-W. Lin and K.~Orginos, {\it {First Calculation of Hyperon Axial Couplings
  from Lattice QCD}},  {\em Phys. Rev.} {\bf D79} (2009) 034507,
  [\href{http://xxx.lanl.gov/abs/0712.1214}{{\tt arXiv:0712.1214}}].

\bibitem{Adamczyk:2014xyw}
{\bf STAR} Collaboration, L.~Adamczyk et~al., {\it {Measurement of longitudinal
  spin asymmetries for weak boson production in polarized proton-proton
  collisions at RHIC}},  {\em Phys. Rev. Lett.} {\bf 113} (2014) 072301,
  [\href{http://xxx.lanl.gov/abs/1404.6880}{{\tt arXiv:1404.6880}}].

\bibitem{Adare:2015gsd}
{\bf PHENIX} Collaboration, A.~Adare et~al., {\it {Measurement of
  parity-violating spin asymmetries in W$^{\pm}$ production at midrapidity in
  longitudinally polarized $p$$+$$p$ collisions}},  {\em Phys. Rev.} {\bf D93}
  (2016), no.~5 051103, [\href{http://xxx.lanl.gov/abs/1504.0745}{{\tt
  arXiv:1504.0745}}].

\bibitem{Anselmino:2008jk}
M.~Anselmino, M.~Boglione, U.~D'Alesio, A.~Kotzinian, F.~Murgia, A.~Prokudin,
  and S.~Melis, {\it {Update on transversity and Collins functions from SIDIS
  and e+ e- data}},  {\em Nucl. Phys. Proc. Suppl.} {\bf 191} (2009) 98--107,
  [\href{http://xxx.lanl.gov/abs/0812.4366}{{\tt arXiv:0812.4366}}].

\bibitem{Radici:2015mwa}
M.~Radici, A.~Courtoy, A.~Bacchetta, and M.~Guagnelli, {\it {Improved
  extraction of valence transversity distributions from inclusive dihadron
  production}},  {\em JHEP} {\bf 05} (2015) 123,
  [\href{http://xxx.lanl.gov/abs/1503.0349}{{\tt arXiv:1503.0349}}].

\bibitem{Radici:2013vra}
M.~Radici, A.~Bacchetta, and A.~Courtoy, {\it {Detecting correlated di-hadron
  pairs: About the extraction of transversity and beyond}},  {\em Int. J. Mod.
  Phys. Conf. Ser.} {\bf 25} (2014) 1460045,
  [\href{http://xxx.lanl.gov/abs/1308.5928}{{\tt arXiv:1308.5928}}].

\bibitem{Kang:2015msa}
Z.-B. Kang, A.~Prokudin, P.~Sun, and F.~Yuan, {\it {Extraction of Quark
  Transversity Distribution and Collins Fragmentation Functions with QCD
  Evolution}},  {\em Phys. Rev.} {\bf D93} (2016), no.~1 014009,
  [\href{http://xxx.lanl.gov/abs/1505.0558}{{\tt arXiv:1505.0558}}].

\bibitem{Schweitzer:2001sr}
P.~Schweitzer, D.~Urbano, M.~V. Polyakov, C.~Weiss, P.~V. Pobylitsa, and
  K.~Goeke, {\it {Transversity distributions in the nucleon in the large N(c)
  limit}},  {\em Phys. Rev.} {\bf D64} (2001) 034013,
  [\href{http://xxx.lanl.gov/abs/hep-ph/0101300}{{\tt hep-ph/0101300}}].

\bibitem{Kang:2014zza}
Z.-B. Kang, A.~Prokudin, P.~Sun, and F.~Yuan, {\it {Nucleon tensor charge from
  Collins azimuthal asymmetry measurements}},  {\em Phys. Rev.} {\bf D91}
  (2015) 071501, [\href{http://xxx.lanl.gov/abs/1410.4877}{{\tt
  arXiv:1410.4877}}].

\bibitem{Bhattacharya:2015wna}
{\bf PNDME} Collaboration, T.~Bhattacharya, V.~Cirigliano, S.~Cohen, R.~Gupta,
  A.~Joseph, H.-W. Lin, and B.~Yoon, {\it {Iso-vector and Iso-scalar Tensor
  Charges of the Nucleon from Lattice QCD}},  {\em Phys. Rev.} {\bf D92}
  (2015), no.~9 094511, [\href{http://xxx.lanl.gov/abs/1506.0641}{{\tt
  arXiv:1506.0641}}].

\bibitem{Bali:2016lva}
G.~S. Bali, B.~Lang, B.~U. Musch, and A.~Schäfer, {\it {Novel quark smearing
  for hadrons with high momenta in lattice QCD}},  {\em Phys. Rev.} {\bf D93}
  (2016), no.~9 094515, [\href{http://xxx.lanl.gov/abs/1602.0552}{{\tt
  arXiv:1602.0552}}].

\bibitem{Alexandrou:2016jqi}
C.~Alexandrou, K.~Cichy, M.~Constantinou, K.~Hadjiyiannakou, K.~Jansen,
  F.~Steffens, and C.~Wiese, {\it {New Lattice Results for Parton
  Distributions}},  \href{http://xxx.lanl.gov/abs/1610.0368}{{\tt
  arXiv:1610.0368}}.

\bibitem{Ishikawa:2016znu}
T.~Ishikawa, Y.-Q. Ma, J.-W. Qiu, and S.~Yoshida, {\it {Practical quasi parton
  distribution functions}},  \href{http://xxx.lanl.gov/abs/1609.0201}{{\tt
  arXiv:1609.0201}}.

\bibitem{Chen:2016fxx}
J.-W. Chen, X.~Ji, and J.-H. Zhang, {\it {Improved quasi parton distribution
  through Wilson line renormalization}},
  \href{http://xxx.lanl.gov/abs/1609.0810}{{\tt arXiv:1609.0810}}.

\bibitem{Edwards:2004sx}
{\bf SciDAC, LHPC, UKQCD} Collaboration, R.~G. Edwards and B.~Joo, {\it {The
  Chroma software system for lattice QCD}},  {\em Nucl. Phys. Proc. Suppl.}
  {\bf 140} (2005) 832, [\href{http://xxx.lanl.gov/abs/hep-lat/0409003}{{\tt
  hep-lat/0409003}}]. [,832(2004)].

\end{thebibliography}
\providecommand{\href}[2]{#2}\begingroup\raggedright\endgroup

\end{document}